\documentclass[a4paper,twoside]{article}

\usepackage{epsfig}
\usepackage{subfigure}
\usepackage{calc}
\usepackage{amssymb}
\usepackage{amstext}
\usepackage{amsmath}
\usepackage{amsthm}
\usepackage{multicol}
\usepackage{pslatex}
\usepackage{apalike}

\usepackage{graphicx}
\usepackage[utf8]{inputenc}  
\usepackage{url}
\usepackage{multirow}
\usepackage[usenames]{color}
\usepackage{pgfplots}
\usepackage{xcolor}

\usepackage{SCITEPRESS}     

\begin{document} 

\title{Efficient Social Network Multilingual Classification using\\Character, POS $n$-grams and Dynamic Normalization}
\author{
	\authorname{
	Carlos-Emiliano Gonz\'alez-Gallardo\sup{1}, Juan-Manuel Torres-Moreno\sup{1,2}, Azucena Montes Rend\'on\sup{3} and Gerardo Sierra\sup{4}
    }
	\affiliation{
    	\sup{1}Laboratoire Informatique d'Avignon, Universit\'e d'Avignon et des Pays de Vaucluse, Avignon, France
    }
	\affiliation{
    	\sup{2}\'Ecole Polytechnique de Montr\'eal, Montr\'eal, Canada
    }
	\affiliation{
    	\sup{3}Centro Nacional de Investigación y Desarrollo Tecnológico, Cuernavaca, M\'exico
    }      
	\affiliation{
    	\sup{4}GIL-Instituto de Ingenier\'ia, Universidad Nacional Aut\'onoma de M\'exico, Ciudad de M\'exico, M\'exico
    }   
\email{
	carlos.gonzalez-gallardo@alumni.univ-avignon.fr, juan-manuel.torres@univ-avignon.fr, \\amr@cenidet.edu.mx, gsierram@iingen.unam.mx
	}
}
\keywords{Text Mining, Machine Learning, Classification, $n$-grams, POS,  Blogs, Tweets, Social Network.}

\abstract{In this paper we describe a dynamic normalization process applied to social network multilingual documents (Facebook and Twitter) to improve the performance of the Author profiling task for short texts. After the normalization process, $n$-grams of characters and $n$-grams of POS tags are obtained to extract all the possible stylistic information encoded in the documents (emoticons, character flooding, capital letters, references to other users, hyperlinks, hashtags, etc.). Experiments with SVM showed up to 90\% of performance.}

\onecolumn \maketitle \normalsize \vfill

\section{\uppercase{Introduction}}
\label{sec:introduction}

\noindent Social networks are a resource that millions of people use to support their relationships with friends and family \cite{chang2015age}. In June 2016, Facebook\footnote{Facebook website: \url{http://www.facebook.com}} reported to have approximately 1.13 billion active users per day; while in June, Twitter\footnote{Twitter website: \url{http://www.twitter.com}} reported an average of 313 million active users per month. 

Due to the large amount of information that is continuously produced in social networks, to identify certain individual characteristics of users has been a problem of growing importance for different areas like forensics, security and marketing \cite{rangel2015overview}; problem that Author profiling aims from an automatic classification approach with the premise that depending of the individual characteristics of each person; such as age, gender or personality, the way of communicating will be different.

Author profiling is traditionally applied in texts like literature, documentaries or essays \cite{argamon2003gender,argamon2009automatically}; these kind of texts have a relative long length \cite{peersman2011predicting} and a standard language. By the other hand, documents produced by social networks users have some characteristics that differ from regular texts and prevents that they can be analyzed in a similar way \cite{peersman2011predicting}. Some characteristics that social networks texts share are their length (significantly shorter that traditional texts) \cite{peersman2011predicting}, the large number of misspellings, the use of non-standard capitalization and punctuation.

Social networks like Twitter have their own rules and features that users use to express themselves and communicate with each other.
It is possible to take advantage of these rules and extract a greater amount of stylistic information. \cite{Gimpel:2011:PTT:2002736.2002747} introduce this idea to create a Part of Speech (POS) tagger for Twitter.
In our case, we chose to perform a dynamic context-dependent normalization.
This normalization allows to group those elements that are capable of providing stylistic information regardless of its lexical variability; this phase helps to improve the performance of the classification process.
In \cite{gonzalez2016perfilado} we presented some preliminary results of this work, that we will extend with more experiments and further description.

The paper is organized as follows:
in section \ref{sec:genderageinden} we provide an overview of Author profiling applied to age, gender and personality traits prediction. In section \ref{sec:ngramascharypos}, a brief presentation of character and POS $n$-grams is presented. In section \ref{sec:nddc} we detail the methodology used in the dynamic context-dependent normalization.
Section \ref{sec:dataset} presents the datasets used in the study.
The learning model is detailed in section \ref{sec:modelo}.
The different experiments and results are presented in section \ref{sec:results}.
Finally, in section \ref{sec:conclusions} we present the conclusions and some future job prospects.

\section{\uppercase{Age, gender and personality traits prediction}}
\label{sec:genderageinden}

\noindent \cite{koppel2002automatically} performed a research with a corpus made of 566 documents of the \textit{British National Corpus (BNC)\footnote{BNC website \url{http://www.natcorp.ox.ac.uk/}}} in which they identified automatically the gender of the author using function words and POS n-grams, obtaining a precision of 80\% with a linear separator.

\cite{doyle2005automatic} constructed a simple classifier to predict gender in a collection of 495 essays of the \textit{BAWE} corpus\footnote{BAWE corpus website: \url{http://www2.warwick.ac.uk/fac/soc/al/research/collect/bawe/}}. The classifier consisted in measuring the distance between the text to classify and two different profiles (man and woman), being the smallest one the class that was assign to the text. A precision of 81\% was reported using character, word and POS n-grams as features. 

In \cite{peersman2011predicting}, the authors collected 1.5 millions of samples of the Netlog social network to predict the age and gender of the authors. They used a model created with support vector machines (SVM) \cite{vapnik} taking into account n-grams of words and characters as features. The reported precision was of 64.2\%.

\cite{DBLP:conf/clef/CarmonaLMPE15} predicted age, gender and personality traits of tweets authors using a high lever representation of features. This representation is composed of discriminatory features (Second Order Atribbutes) \cite{lopez2013montes} and descriptive features (Latent Semantic Analysis) \cite{wiemer2004latent}. For age and gender prediction in Spanish tweets they obtained a precision of 77.27\% and for personality traits a RMSE value of 0.1297.

In \cite{DBLP:conf/clef/GrivasKG15}, a classification model with SVM was used to predict age, gender and personality traits of Spanish tweets while a SVR model was used to predict personality traits of tweets authors. In this approach two types of feature groups where considered: structural and stylometric.
Structural features were extracted form the unprocessed tweets while stylometric features were obtained after html tags, urls, hashtags and references to other users are removed. A precision of 72.73\% was reported for age and gender prediction, while a RMSE value of 0.1495 was obtained for personality traits prediction.

\section{\uppercase{Character and POS $n$-grams}}
\label{sec:ngramascharypos}

\noindent $n$-grams are sequences of elements of a selected textual information unit \cite{manning:99} that allow the extraction of content and stylistic features from text; features that can be used in tasks such as Automatic Summarization, Automatic Translation and Text Classification.

The information unit to use changes depending on the task and the type of features that will be extracted.
For example, in Automatic Translation and Summarization is common to use word and phrase $n$-grams \cite{torres:2014,giannakopoulos2008testing,Koehn:2010:SMT:1734086}.
Within Text Classification, for Plagiarism Detection, Author Identification and Author Profiling; character, word and POS $n$-grams are used \cite{doyle2005automatic,stamatatos2015overview,oberreuter2013text}.

The information units selected in this research are characters and POS tags.
With the character $n$-grams the goal is to extract as many stylistic elements as possible: characters frequency, use of suffixes (gender, number, tense, diminutives, superlatives, etc.), use of punctuation, use of emoticons, etc \cite{Stamatatos2006,Stamatatos2009}.

POS $n$-grams provide information about the structure of the text: frequency of grammatical elements, diversity of grammatical structures and the interaction between grammatical elements.
The POS tags were obtained using the Freeling\footnote{Freeling website: \url{http://nlp.lsi.upc.edu/freeling/node/1}} POS tagger, which follows the POS tags proposed by EAGLES\footnote{EAGLES website: \url{http://www.ilc.cnr.it/EAGLES96/annotate/node9.html}}.
To fully control the standardization process and make it independent of a detector of names, we preferred to perform a specific normalization for both datasets instead of using the Freeling functionalities \cite{padro12}.

The POS tags provided by Freeling have several levels of detail that provide insight into the different attributes of a grammatical category depending of the language that is being analyzed.
In our case we used only the first level of detail that refers to the category itself. The example in Table \ref{table:posespecializacion} shows the Spanish word \textit{"especializaci\'on"} (specialization), all its attributes and the selected tag.

\begin{table}[h]
\caption{Part-of-speech tagging of the Spanish word:  \textit{especializaci\'on}}
\label{table:posespecializacion}
\centering
    \begin{tabular}{|c|c|c|c|}
    \hline
    \multicolumn{4}{|c|}{Word: \textit{"especializaci\'on"}} \\ \hline
    Attribute & Code & Value & Tag\\ \hline
    Category & N & Noun&\\
    Type     & C & Common &\\
    Gender    & F & Female &\\
    Number & S & Singular & \textit{N}\\
    Case     & 0 & - &\\
    Semantic    & 0 & - &\\
    Gender       & 0 & -   &\\
    Grade   & 0 & - &\\
    \hline
    \end{tabular}
\end{table}

\section{\uppercase{Dynamic Context-Dependent Normalization}}
\label{sec:nddc}

\noindent The freedom that social networks users have to encode their messages derives in a varied lexicon. To minimize this variations it is necessary to normalize those elements that are able to provide stylistic information regardless of its lexical variability (references to other users, hyperlinks, hashtags, emoticons, etc). This is what we call Dynamic Context-dependent Normalization which is separated in two phases: Text normalization and POS relabeling.
\begin{itemize}
	\item Text normalization\\
    The goal of this phase is to avoid the lexical variability that is present when a user tends to do certain actions in their texts; for example tagging another user or creating a link to a website.
    In the Twitter case, references to other users are defined as

    \begin{small}
    \begin{verbatim}
    @{username}
    \end{verbatim}
    \end{small}

    The amount of values that can be assigned to the label \textit{username} is potentially infinite (depending on the number of users available to the social network).
    To avoid such variability, the Dynamic Context-dependent Normalization standardize this element in order to highlight the intention: make a reference to a user. So, the tweet 

    \begin{small}
    \begin{verbatim}
    I was just watching ``update 10.''
    @MKBHD http://t.co/P9Dn7t8zSl
    \end{verbatim}
    \end{small}
    
    will be normalized to

	\begin{small}
    \begin{verbatim}
    I was just watching ``update 10.''
    @us http://t.co/P9Dn7t8zSl
    \end{verbatim}
    \end{small}    

Hyperlinks have a similar behavior: the number of links to Internet sites is also potentially infinite.
    The important fact is that a reference to an external site is being implemented; so all text strings that match the pattern:

    \begin{small}
    \begin{verbatim}
    http[s]://{external_site}
    \end{verbatim}
    \end{small}

    \noindent are normalized. Following with the previous example, the tweet
    
	\begin{small}
    \begin{verbatim}
    I was just watching ``update 10.''
    @us http://t.co/P9Dn7t8zSl
    \end{verbatim}
    \end{small} 
    
    will be normalized to 

	\begin{small}
    \begin{verbatim}
    I was just watching ``update 10.''
    @us htt
    \end{verbatim}
    \end{small}   

    \item POS relabeling\\
    All these lexical variations provide important grammatical information that must be preserved, but conventional POS taggers are unable to maintain.
Therefore, it is necessary to relabel certain elements to keep their interaction with the rest of the POS tags of the text.

	Using Freeling to tag the previous example and taking into account only the categories of the POS tag, the sequence obtained is

	\begin{small}
    \begin{verbatim}
	P V R V N Z F N N
    \end{verbatim}
    \end{small}   

As it can be seen, the reference to the user and the link to the site is lost and it is no possible to know how those elements interact with the rest of the tags.
To overcome this limitation, references to other users, hyperlinks and hashtags are relabeled so that they have their own special tag dealing to the next sequence:

	\begin{small}
    \begin{verbatim}
	P V R V N Z REF@USERNAME REF#LINK 
    \end{verbatim}
    \end{small}   

\end{itemize}

\begin{figure}[!h]
\centering
    \includegraphics[height=7cm,keepaspectratio]{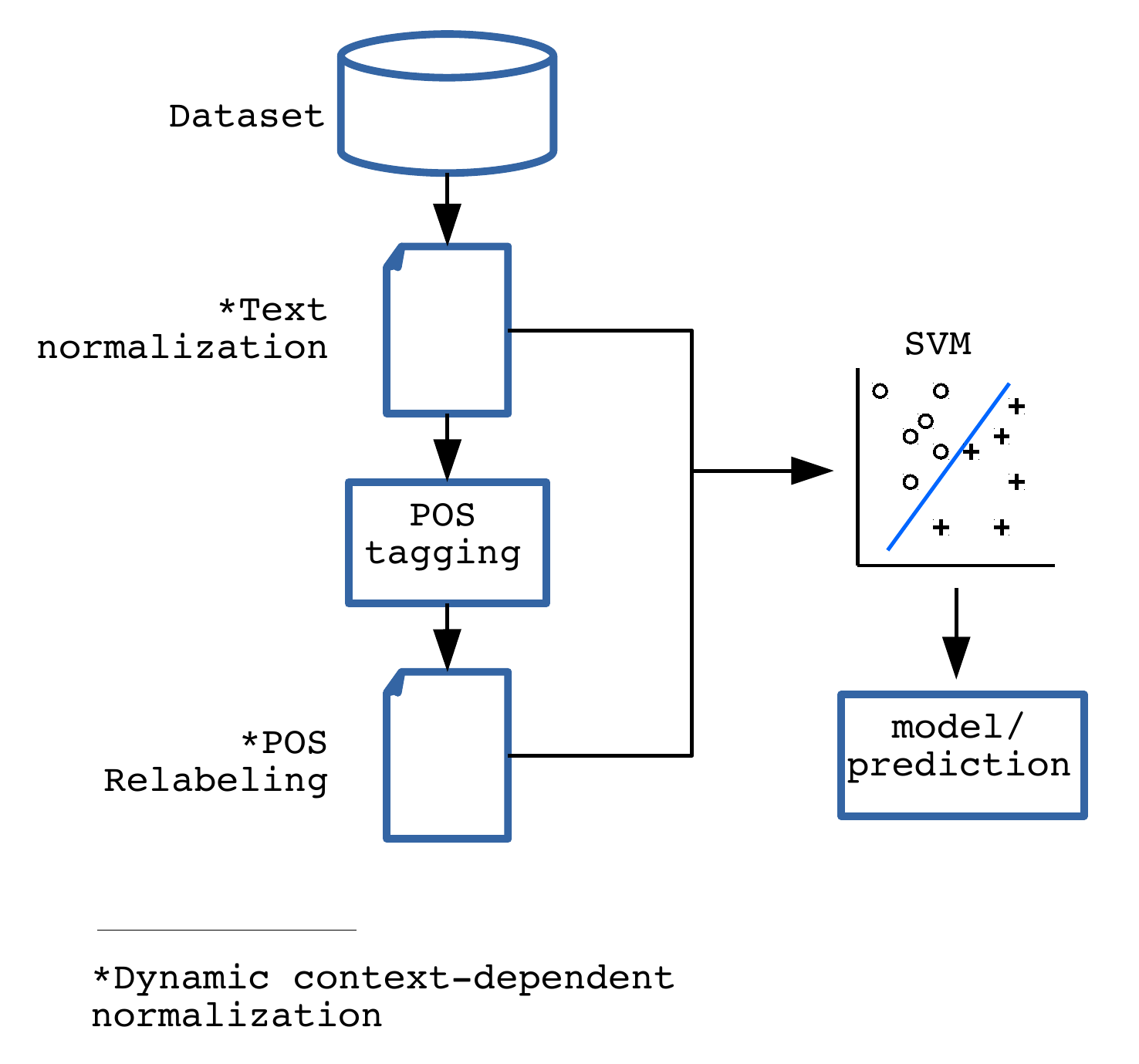}
  \caption{\label{fig:confmax1} General architecture of classification system.}
\end{figure}

\noindent Figure~ \ref{fig:confmax1} shows a general architecture of the system.

\section{\uppercase{Datasets}}
\label{sec:dataset}

\noindent In order to test various contexts, we used corpora from two social networks: Twitter and Facebook.

The multilingual corpus PAN-CLEF2015 (Twitter) is labeled by gender, age and personality traits; whereas the multilingual corpus PAN-CLEF2016 (Twitter) is just labeled by gender and age.
The Mexican Spanish corpus ``Comments of Mexico City through time'' (Facebook) is only labeled by gender.

\subsection{PAN-CLEF2015}

\noindent The PAN-CLEF2015 corpus\footnote{PAN challenge's website: ~\url{http://pan.webis.de/}} \cite{Rangel2015} is conformed by 324 samples distributed in four languages:
Spanish, English, Italian and Dutch. A larger description of the corpus can be seen in \cite{gonzalez2016perfilado}.

\subsection{PAN-CLEF2016\_s}

\noindent The multilingual PAN-CLEF2016\_s corpus is a set of the training corpus for the Author Profiling task of PAN 2016.\footnote{The corpus is downloadable at the Website: \url{http://www.uni-weimar.de/medien/webis/events/pan-16/pan16-code/pan16-author-profiling-twitter-downloader.zip}} 
It is conformed by 935 samples labeled by age and gender which are distributed in three languages:
Spanish, English and Dutch.
For our experiments, only the gender label was considered.

Table \ref{table:PAN-CLEF 2016_distribuciongenero} shows the distribution of gender samples per language.
\begin{table}[h]
    \caption{PAN-CLEF2016\_s, distribution of samples by gender.}
    \label{table:PAN-CLEF 2016_distribuciongenero}
\centering
\begin{tabular}{|l|c|c|c|}
\hline
 & \multicolumn{2}{c|}{Samples}  \\ 
 & Female                & Male   \\ \hline
Spanish                   & 52\%                    & 48\% \\
English                   & 52\%                     & 48\% \\
Dutch                     & 50\%                   & 50\% \\ \hline 
\end{tabular}
\end{table}

\subsection{Comments of Mexico City through time (CCDMX)}

\noindent The CCDMX corpus consists of 5 979 Mexican Spanish comments from the Facebook page \textit{Mexico City through time}\footnote{Website of the blog:~ \url{ http://www.facebook.com/laciudaddemexicoeneltiempo}}.
In this Facebook page, pictures of Mexico City are posted so that people can share their memories and anecdotes.
The average length of each comment is 110 characters.

The CCDMX corpus was manually annotated by bachelor's linguistic students of the Group of Linguistic Engineering (GIL) of the UNAM in 2014\footnote{Website of the corpus: \url{http://corpus.unam.mx}}. It is only labeled by gender, being slightly higher the number of comments that belong to the ``Men'' class (see table \ref{table:distribucioncorpusCDMX}). 

\begin{table}[h]
  \caption{CCDMX, distribution of samples by gender.}
\label{table:distribucioncorpusCDMX}
\centering
\begin{tabular}{|c|c|c|}
\hline
  & Comments & Percentage\\ \hline
Female & 2573 & 43\%\\
Male   & 3406 & 57\%\\\hline
Total of comments & 5 979 & 100\% \\
\hline
\end{tabular}
\end{table}

\section{\uppercase{Learning Model}}
\label{sec:modelo}

\noindent For the experiments we used support vector machines (SVM) \cite{vapnik}, a classical model of supervised learning, which has proven to be robust and efficient in various NLP tasks.

In particular, to perform experiments we use the Python package \textit{Scikit-learn}\footnote{Scikit-learn is downloadable at the Website: \url{http://scikit-learn.org}}\cite{Pedregosa2011} using a linear kernel (LinearSVC), which produced empirically the best results.

\subsection{Features}
\label{sec:parametros}

\noindent The character and POS $n$-gram windows used were generated with a unit length of 1 to 3.

Thus, for example, the word \textit{``self-defense''} is represented by the following character $n$-grams:\\

\hrule
\noindent \{\textit{s, e, l, f, -, d, e, f, e, n, s, e, \_s, se, el, lf, f-, -d, de, ef, fe, en, ns, se, e\_, \_se, sel, elf, lf-, f-d, -de, def, efe, fen, ens, nse, se\_}\}
\hrule

\vspace{0.5cm}
\noindent And the normalize tweet ``\texttt{@us @us You owe me one, Cam!}'' which POS sequence is ``\texttt{REF@USERNAME REF@USERNAME N V P N F N F}'', is represented by the POS $n$-grams: \\
\vspace{0.5cm}
\hrule
\noindent \{\texttt{REF@USERNAME, REF@USERNAME, N, V, P, N, F, N, F,REF@USERNAME REF@USERNAME, REF@USERNAME N, N V, V P, P N, N F, F N, N F, REF@USERNAME REF@USERNAME N, REF@USERNAME N V, N V P, V P N, P N F, N F N, F N F}\}
\hrule

\vspace{0.2cm}
The best results were obtained using a linear frequency scale in all cases except from the POS $n$-grams for Spanish texts, in which the logarithmic function

\begin{equation*}
  \log_2(1+number\_of\_occurrences) 
\end{equation*}

\begin{figure}[!h]
  \centering
  \includegraphics[width=0.51\textwidth]{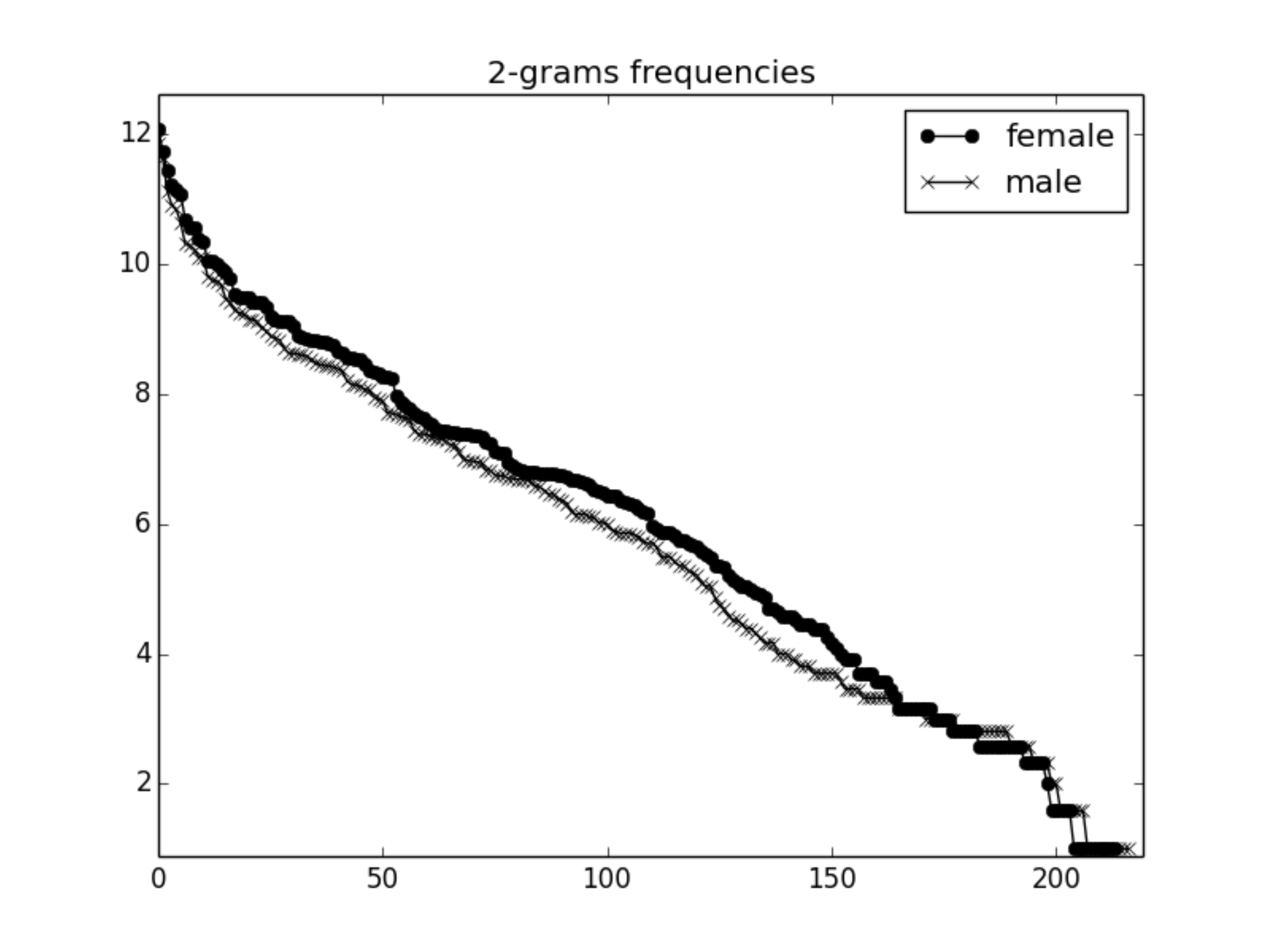} 
  \caption{Linearized frequencies of the Spanish PAN-CLEF2015 corpus}
  \label{fig:2-grams}
 \end{figure}

\vspace{0.2cm}
\noindent is applied. In figure \ref{fig:2-grams} is possible to see the linearized POS 2-grams values of the PAN-CLEF2015 training corpus.

\subsection{Experimental Protocol}

Four experiments were performed with the PAN-CLEF2015 corpus; one for each language.
70\% of the samples was used for training the learning model and the remaining 30\% was used during evaluation time.

Regard to the PAN-CLEF2016\_s corpus,three experiments were performed; one for Dutch, one for English and another one for Spanish.
We used the training model created with the PAN-CLEF2015 corpus and applied it to this corpus.
All the samples available in the PAN-CLEF2016\_s corpus were used as test samples for gender classification.

Three experiments were performed with the CCDMX corpus.
\begin{itemize}
\item First, all the comments were used as test samples using the learning model generated with the Spanish training samples of the PAN-CLEF2015 corpus.

\item For the second experiment, samples of 50 comments were created. Thus 121 samples were tested using the same learning model of the first experiment.

\item Finally, the third experiment was a sum of various micro experiments.
For each micro experiment a different sample size was tested: 1, 2, 4, 8, 16, 24, 32, 41, 50, 57 and 64 comments per sample. 70\% of the samples were used to train the learning model and 30\% to test it. 

\end{itemize}

\section{\uppercase{Results}}
\label{sec:results}

\noindent In order to evaluate the performance of the model in all the datasets, a group of classical measures was implemented.
Accuracy (\textrm{A}), Precision (\textrm{P}), Recall (\textrm{R}) and F-score (\textrm{F}) \cite{manning:99} were used to evaluate gender and age prediction. 
In relation to the personality traits prediction in the PAN-CLEF2015 corpus, the Root Mean Squared Error (RMSE) measure was used.

\subsection{PAN-CLEF2015 Corpus}

Tables \ref{table:CLEF_resultados_genero_espanol} to \ref{table:CLEF_resultados_rasgos_ingles}  show the results obtained from the PAN-CLEF2015 corpus for Spanish and English. 

Evaluation measures reports practically 100\% in gender classification for Italian and Dutch \cite{gonzalez2016perfilado}. This is probably because the number of existing samples for both languages was very small.
We think it would be worth trying with a larger amount of data to validate the results in these two languages.
\begin{itemize}
  \item Spanish. 
    \begin{table}[ht!]
    \caption{PAN-CLEF2015; gender and age results (Spanish).}
    \label{table:CLEF_resultados_genero_espanol}
    \centering	
    \begin{tabular}{|c|c|c|c|c|}
        \hline
    Gender   & P     & R    & F    & A \\ \hline
    Male & 0.929 & 0.867& 0.897& \multirow{2}{*}{0.900}\\
    Female & 0.875 & 0.930 & 0.902&\\\hline 

        \hline
    Age   & P     & R    & F     & A \\ \hline
    18-24 & 0.750 & 1    & 0.857 &\multirow{4}{*}{0.800}\\ 
    25-34 & 0.750 & 0.875& 0.807 &\\
    35-49 & 1     & 0.667& 0.800 &\\
    $>$50 & 1     & 0.500& 0.667 &\\\hline
    \end{tabular}
    \end{table}

    \begin{table}[ht!]
    \caption{PAN-CLEF2015; personality traits results (Spanish).}
    \label{table:CLEF_resultados_rasgos_espanol}
    \centering
    \begin{tabular}{|c|c|}
        \hline
    Trait & RMSE \\ \hline
    E & 0.106\\
    N & 0.128\\
    A & 0.158\\
    C & 0.164\\
    O & 0.138\\\hline
    Mean & 0.139 \\\hline
    \end{tabular}
    \end{table}  
    
  \item English.\newline
    \begin{table}[ht!]
    \caption{PAN-CLEF2015; gender and age results (English).}
    \label{table:CLEF_resultados_genero_ingles}
    \centering
    \begin{tabular}{|c|c|c|c|c|}
        \hline
    Gender   & P     & R    & F    & A \\ \hline
    Male   & 0.826 & 0.826 & 0.826 &\multirow{2}{*}{0.826}\\
    Female & 0.826 & 0.826 & 0.826 &\\\hline
%
    \hline
    Age   & P     & R    & F     & A \\ \hline
    18-24 & 0.895 & 0.944 & 0.919&\multirow{4}{*}{0.848}\\
    25-34 & 0.789 & 0.833 & 0.810 &\\
    35-49 & 0.800 & 0.667 & 0.727 &\\
    $>$50 & 1     & 0.750 & 0.857 &\\\hline
\end{tabular}
\end{table}

    \begin{table}[ht!]
    \caption{PAN-CLEF2015; personality traits results (English).}
    \label{table:CLEF_resultados_rasgos_ingles}
    \centering
    \begin{tabular}{|c|c|}
        \hline
    Trait & RMSE \\ \hline
    E & 0.182\\
    N & 0.182\\
    A & 0.150\\
    C & 0.123\\
    O & 0.162\\\hline
    Mean & 0.160 \\\hline
    \end{tabular}
    \end{table} 
    
    \end{itemize}

\subsection{PAN-CLEF2016\_s Corpus}

\noindent As it can be see in tables \ref{table:CLEF2016_resultados_genero_espanol}  to \ref{table:CLEF2016_resultados_genero_holandes}, the performance of the experiments is low in all three languages.
We think this is caused by the difference that exists in the size of the samples used to train the model (PAN-CLEF2015 corpus) and the size of the PAN-CLEF2016\_s corpus samples. Another possible reason is that the quantity of noise in the PAN-CLEF2016\_s corpus was to much to be handled by the trained models; thus creating mistaken predictions.

\begin{itemize}
  \item Spanish.\newline
    \begin{table}[ht!]
    \caption{PAN-CLEF2016\_s; gender results (Spanish).}
    \label{table:CLEF2016_resultados_genero_espanol}
    \centering
    \begin{tabular}{|c|c|c|c|c|}
        \hline
    Gender& P     & R & F    & A \\ \hline
    Male   & 0.494 & 0.967 & 0.654&\multirow{2}{*}{0.505}\\
    Female & 0.700     & 0.071 & 0.130 &\\\hline
    \end{tabular}
    \end{table}

  \item English.\newline
    \begin{table}[ht!]
    \caption{PAN-CLEF2016\_s; gender results (English).}
    \label{table:CLEF2016_resultados_genero_ingles}
    \centering
    \begin{tabular}{|c|c|c|c|c|}
        \hline
    Gender& P     & R & F    & A \\ \hline
    Male   & 0.578 & 0.531 & 0.554&\multirow{2}{*}{0.587}\\
    Female & 0.594     & 0.638 & 0.615 &\\\hline
    \end{tabular}
    \end{table}

  \item Dutch.\newline
	Number of samples: 382.
    \begin{table}[ht!]
    \caption{PAN-CLEF2016\_s; gender results (Dutch).}
    \label{table:CLEF2016_resultados_genero_holandes}
    \centering
    \begin{tabular}{|c|c|c|c|c|}
        \hline
    Gender& P     & R & F    & A \\ \hline
    Male   & 0.774 & 0.251 & 0.379&\multirow{2}{*}{0.589}\\
    Female & 0.553     & 0.927 & 0.693 &\\\hline
    \end{tabular}
    \end{table}
\end{itemize}

\subsection{CCDMX Corpus}

\noindent The first experiment (E1) performed with this corpus aimed to discover how much impact had the difference between the training and test samples. 
The training phase was done with 70\% of the samples from the PAN~-CLEF2015 corpus (Spanish). Remember that a sample of this corpus is a group of about 100 tweets.

Table \ref{table:CDMX_resultados_e1} shows the results of the 5 979 samples that were tested.

\begin{table}[ht!]
\caption{CCDMX, E1 results.}
\label{table:CDMX_resultados_e1}
\centering
\begin{tabular}{|l|c|c|c|c|}
    \hline
        & P     & R     & F     & A \\ \hline
Male & 0.598 & 0.631 & 0.614 & \multirow{2}{*}{0.549}\\
Female & 0.474 & 0.439 & 0.456 &\\\hline
\end{tabular}
\end{table}

In the second experiment (E2) we chose to generate samples of 50 comments, size that represent a reasonable compromise between number of samples and number of characters per sample (about 5K characters).

A total of 121 samples were tested with the learning model of E1. The results are slightly better that E1 but the domain difference seems to affect greatly the system performance (Table \ref{table:CDMX_resultados_e2}).

\begin{table}[ht!]
\caption{CCDMX; E2 results.}
\label{table:CDMX_resultados_e2}
\centering
\begin{tabular}{|l|c|c|c|c|}
    \hline
        & P     & R     & F     & A \\ \hline
Male & 0.657 & 0.942 & 0.774 &\multirow{2}{*}{0.686}\\
Female & 0.818 & 0.346 & 0.486 &\\\hline
\end{tabular}
\end{table}

A third experiment (E3) was done with this corpus; eleven micro experiments with different number of comments per sample were performed: 1, 2, 4, 8, 16, 24, 32, 41, 50, 57, 64. This was done to measure the impact of the sample size variation in the performance of the algorithm.

\begin{figure}[!h]
  \centering
  \includegraphics[width=0.51\textwidth]{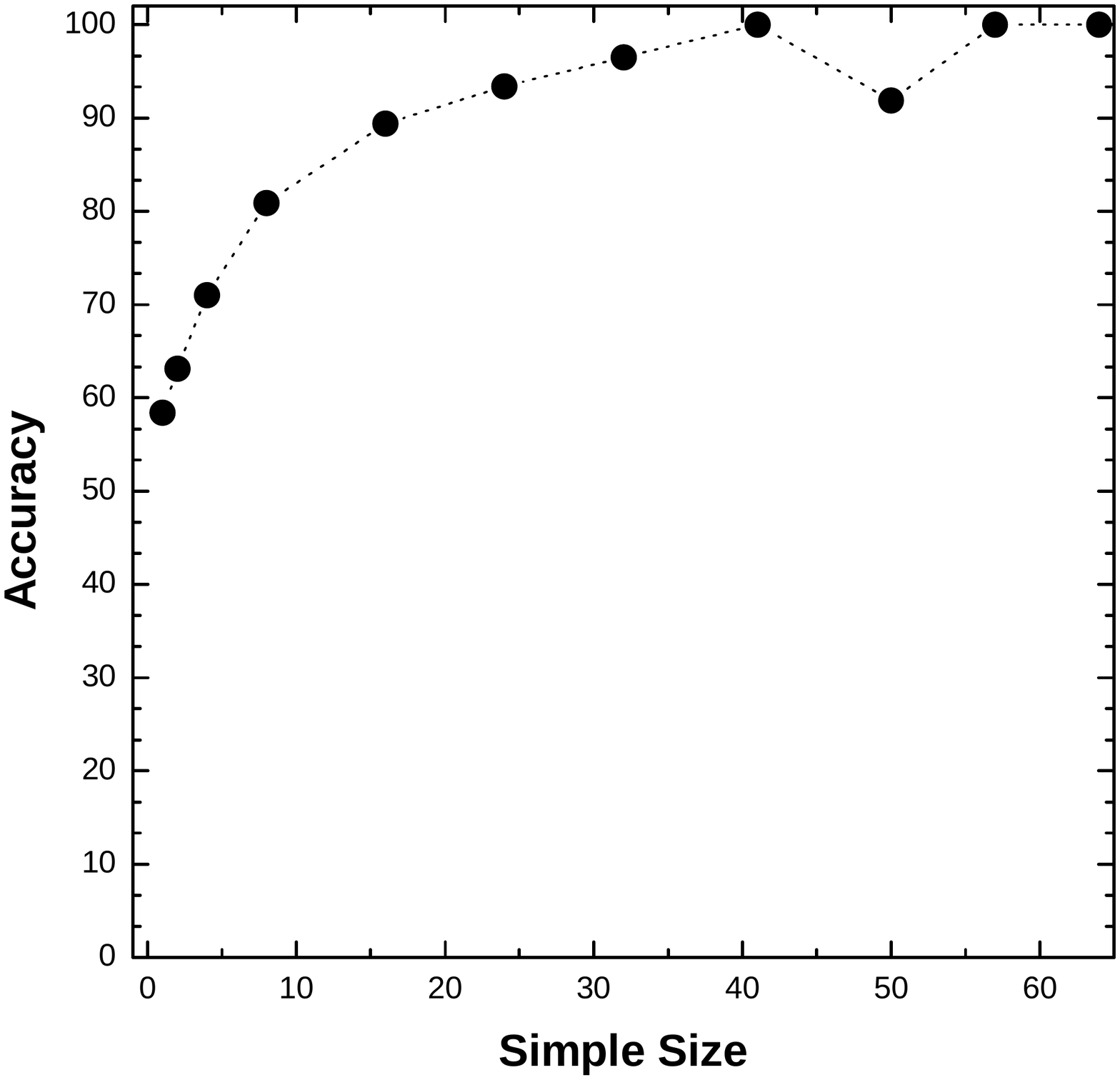} 
  \caption{CCDMX; Accuracy}
  \label{fig:accuracy}
 \end{figure}

\begin{figure}[!h]
  \centering
  \includegraphics[width=0.51\textwidth]{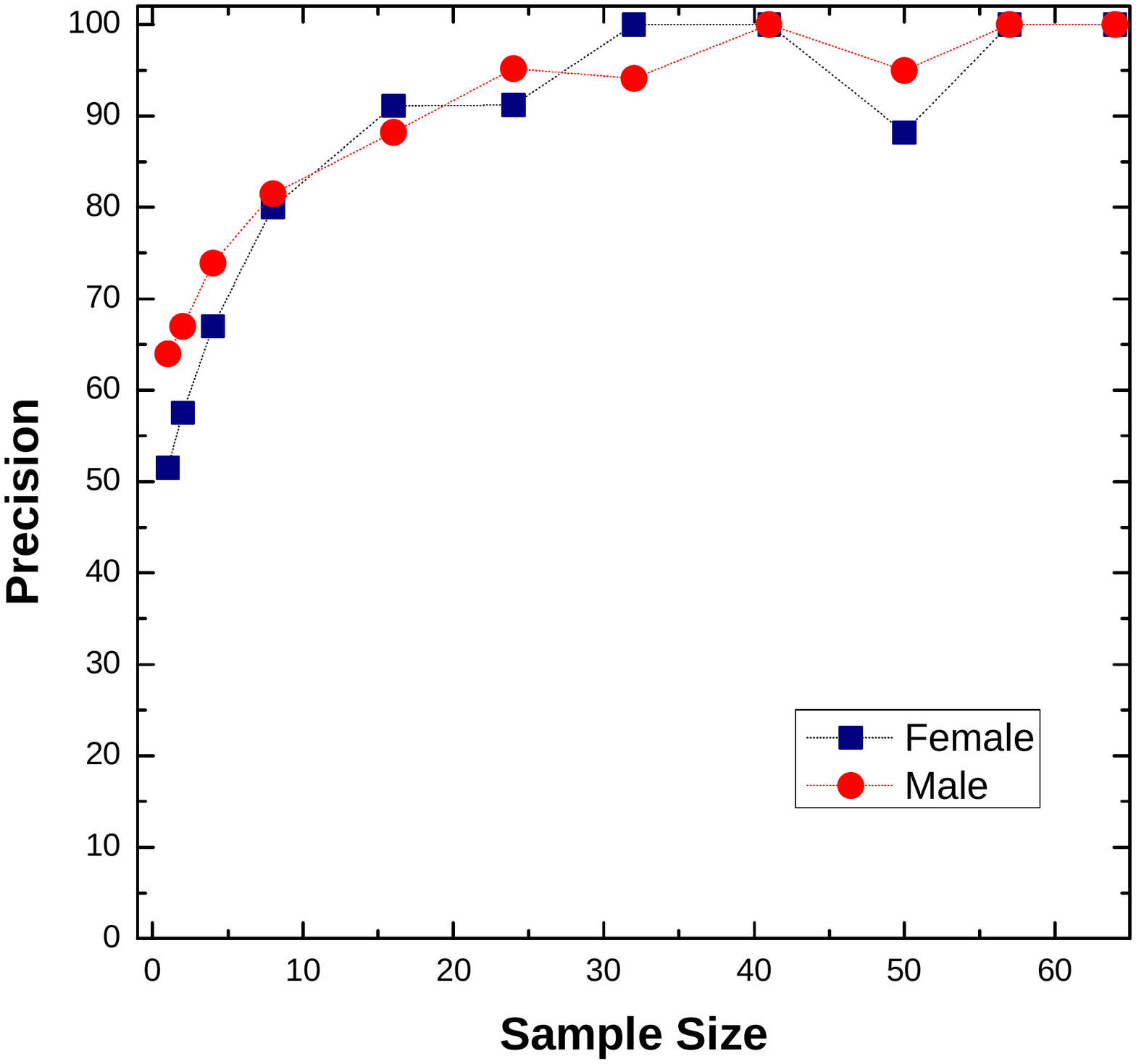} 
  \caption{CCDMX; Precision}
  \label{fig:precision}
 \end{figure}

\begin{figure}[!h]
  \centering
  \includegraphics[width=0.51\textwidth]{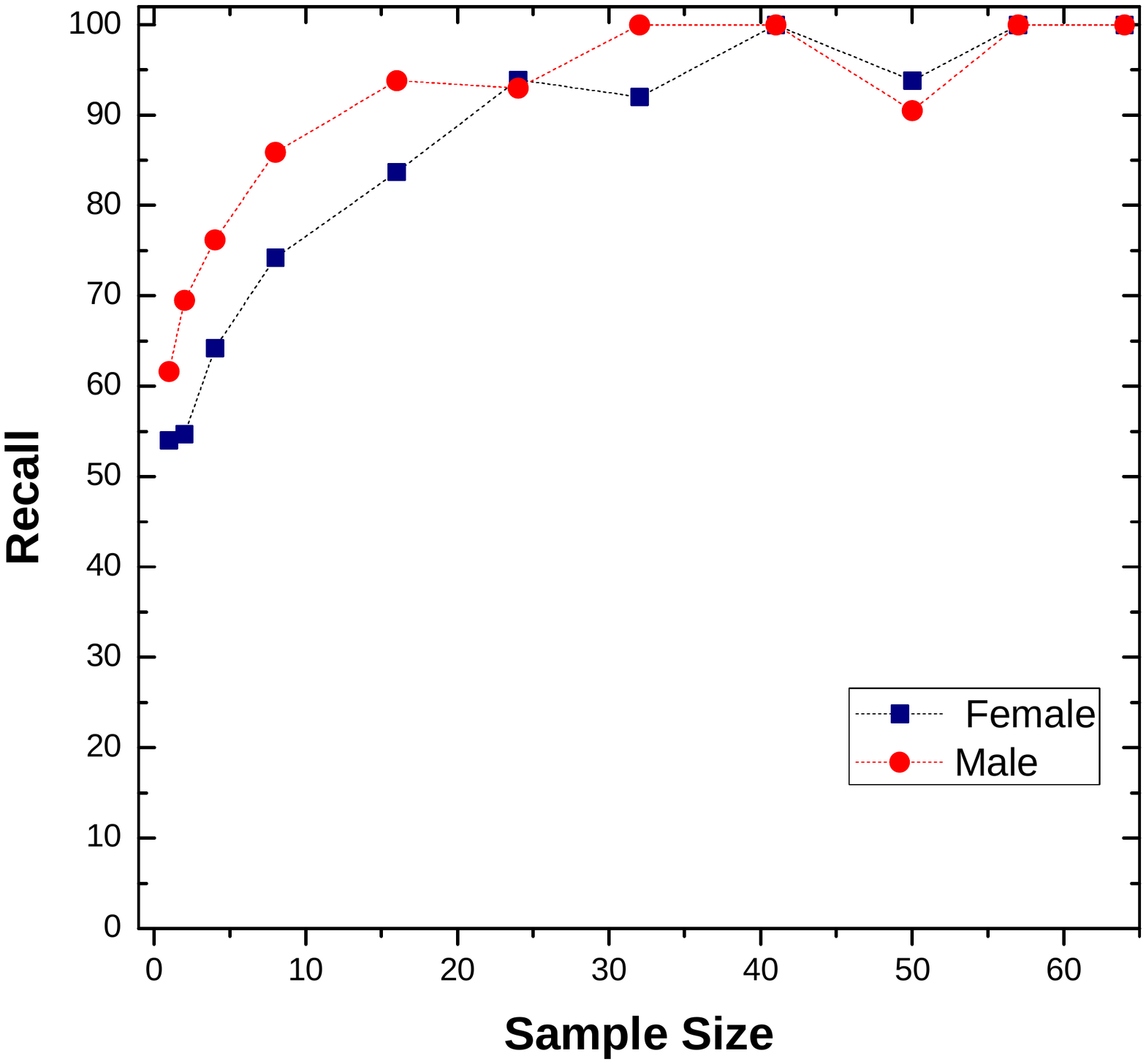} 
  \caption{CCDMX; Recall}
  \label{fig:recall}
 \end{figure}

\begin{figure}[!h]
  \centering
  \includegraphics[width=0.51\textwidth]{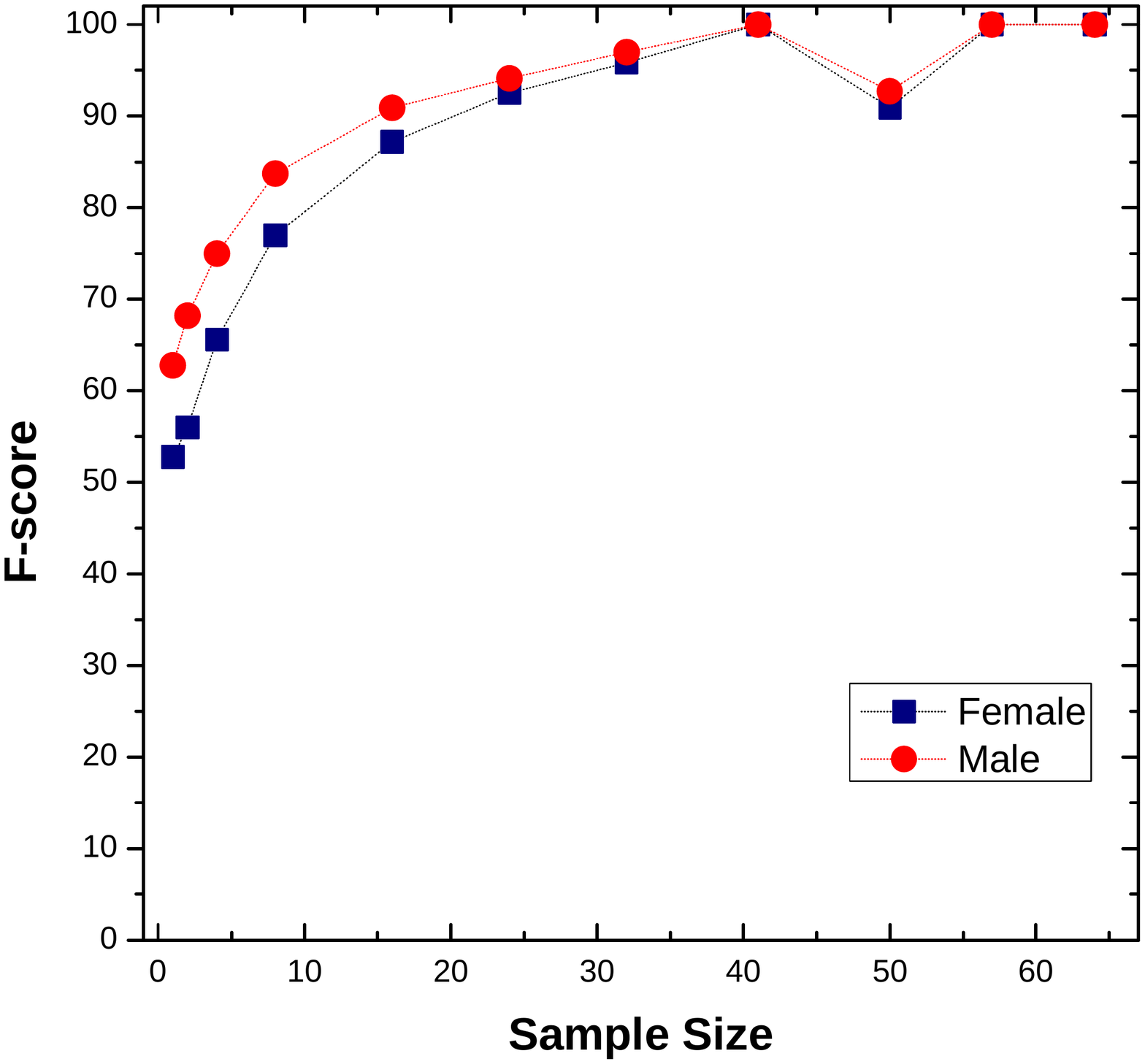} 
  \caption{CCDMX; F-score}
  \label{fig:fscore}
 \end{figure}

Figures \ref{fig:accuracy} to \ref{fig:fscore} show the
performance of E3 (accuracy, precision, recall and F-score). As it can be seen, something curious happens when the sample is of 50 comments length. At that point the general performance drops about 8\%. 
This is due to an statistical effect that the data shows with the residual sample. For both male and female, a residual sample of 22 comments is present. It is necessary to mention that this effect does not affect the general results of E3, which shows to be consistent with the rest of the experiments.

\section{\uppercase{Conclusions and Future Work}}
\label{sec:conclusions}

\noindent Character and POS $n$-grams have shown to be a useful resource for the extraction of stylistic features in short texts. 
With character $n$-grams it was possible to extract emoticons, punctuation exaggeration (character flooding), use of capital letters and different kinds of emotional information encoded in tweets and comments.
The interaction between the special elements of the social networks, like references to users, hashtags or links, and the rest of the POS tags were captured with the POS $n$-grams after. Also, for Spanish and English it was possible to capture the most representative series of two and three grammatical elements; for the Italian and Dutch we were able to capture the most common grammatical elements.
The Dynamic Context-dependent Normalization showed to be effective in the different domains (Facebook and Twitter), but also showed to have problems when a cross domain evaluation was performed.
The poor results obtained in E1 reflect that it is important to keep the ratio size between train and test samples. E2 results show that the change of domain greatly affects the prediction capacity of the system; fact that E3 and the results obtained with the PAN-CLEF2016\_s corpus reinforce.
A proposal for future work is to repeat E3 applying cross validation. This will show a better distribution of the training and test samples eliminating the possibility of a biased result. Also, this probably balance the 50 comments samples and improves its performance.
An interesting development for future work would be to apply the algorithm presented in this article to comments of multimedia sources to separate in an automatic way the different sectors of opinion makers.

\section*{\uppercase{Acknowledgements}}
\noindent This project was partially financed by the project: CONACyT-M\'exico No. 215179 \textit{Caracterizaci\'on de huellas textuales para el an\'alisis forense}. We also appreciate the financing of the european project CHISTERA CALL - ANR: \textit{Access Multilingual Information opinionS (AMIS)}, (France - Europe).

\bibliographystyle{apalike}
{\small
\bibliography{biblio}}

\vfill

\end{document}